\documentclass[conference]{IEEEtran}

\ifCLASSINFOpdf
\else
\fi
\usepackage{hyperref}
\usepackage{graphicx}
\usepackage{cite}
\usepackage{setspace}
\usepackage{cite}
\usepackage{color}
\usepackage{amsmath}
\usepackage{amsfonts}
\usepackage{amssymb}
\usepackage{amsthm}

\newtheorem{thm}{Theorem}

\title{On LDPC Codes for Gaussian Interference Channels}

\author{
    \IEEEauthorblockN{Shahrouz Sharifi\IEEEauthorrefmark{1}, A.~Korhan~Tanc\IEEEauthorrefmark{2}, Tolga~M.~Duman\IEEEauthorrefmark{1}\IEEEauthorrefmark{3}\\
    \IEEEauthorblockA{\IEEEauthorrefmark{1}School of  ECEE, Arizona State Univ., Tempe, AZ, USA, email: sh.sharifi@asu.edu.\\
	\IEEEauthorblockA{\IEEEauthorrefmark{2}Dept. of EEE, Kirklareli Univ., Kayali Campus, 39100, Kirklareli, Turkey, email: korhan.tanc@kirklareli.edu.tr.\\
    \IEEEauthorblockA{\IEEEauthorrefmark{3}Dept. of EEE, Bilkent Univ., Bilkent, 06800, Ankara, Turkey, email: duman@ee.bilkent.edu.tr.}}
\vspace*{-1cm}}}}

\begin{document}

\maketitle 




\begin{abstract}
In this paper, we focus on the two-user Gaussian interference channel~(GIC), and  study the Han-Kobayashi~(HK) coding/decoding strategy with the objective of designing low-density parity-check~(LDPC) codes. A code optimization algorithm is proposed which adopts a~\emph{random perturbation technique} via tracking the average mutual information. The degree distribution optimization and convergence threshold computation are carried out for strong and weak interference channels, employing binary phase-shift keying~(BPSK). Under strong interference, it is observed that optimized codes operate close to the capacity boundary. For the case of weak interference, it is shown that via the newly designed codes, a nontrivial rate pair is achievable, which is not attainable by single user codes with time-sharing. Performance of the designed LDPC codes are also studied for finite block lengths through simulations of specific codes picked from the optimized degree distributions. 
\end{abstract}
\section{Introduction}
Full characterization of the capacity region of the two-user GIC is an open problem for the general case, and only inner and outer bounds are available in the literature. The best reported achievable rate region to date is due to Han and Kobayashi~(HK)~\cite{Kobayashi1981}. Despite the superiority of the rate region, there is no work on exploring explicit and implementable channel codes adopting this coding and decoding scheme in the current literature. With this motivation, in this paper, we study the design and performance of low-density parity-check~(LDPC) over GICs utilizing the HK strategy. 

In the current literature, LDPC codes have been successfully optimized for multi-user channels, where promising results are obtained for the two-user equal gain multiple-access-channel~(MAC)\cite{Roumy2007}, Gaussian broadcast channel~\cite{Berlin2005}, relay channel~\cite{Hu2007}, and symmetric two-user GIC~\cite{Bennatan2013} where identical distributions for both messages are used without employing the HK coding/decoding strategy. In this paper, we investigate the performance of irregular LDPC codes over the two-user GIC employing the HK coding/decoding scheme with fixed channel gains and finite constellations.  In the proposed HK scheme information of each transmitter is split into private and public parts which are encoded using separate LDPC codes. The encoded bits are mapped to specific constellation points and the resulting signals are superimposed to generate the transmitted signal. At the receiver side, public messages and private message of the intended user are decoded concurrently utilizing an iterative joint decoder. It is shown that the proposed joint decoder enjoys a symmetry property of the exchanged soft information which plays a key role in simplifying the mutual information calculations. 

A code optimization algorithm is proposed based on random perturbations. The algorithm can be considered as a specific instance of differential evolution technique of~\cite{Storn1997}, which is a robust and effective method. The optimization steps through random perturbations starting from \emph{admissible degree distributions}~\cite{Bennatan2013}. The convergence of ensembles is verified by tracking the mutual information evolution utilizing Monte-Carlo simulations.

Computation of HK achievable rate region is prohibitively difficult since full characterization of the rate region requires optimization over numerous random variables with large cardinalities. Therefore, in this paper, instead of computing the entire region, a subregion~\cite{Sharifi2014} is computed with a smaller complexity, where a finite number of power allocations are considered and no time sharing~(TS) is utilized. 

Having implemented the HK strategy, we carry out the code optimization for the two-user GIC through examples considering strong and weak interference. For comparison purposes we will use single user codes with TS implemented with two different power constraints. The first one is naive TS motivated by practical limitations on the power amplifiers, for which we have individual power constraints for the two users for each transmitted symbol. The second one is non-naive TS where the users can ``pool'' their power resources and increase their individual power levels for certain fraction of the transmission while keeping the total average power over the entire codeword under a certain value~\cite{ElGamal2011}.
Promising results are obtained under strong interference, where a rate pair very close to the capacity boundary is achieved. Under weak interference, it is demonstrated that a non-trivial point, which is not achievable with the point-to-point~(p2p) code used with TS, is attainable. In our examples, we also evaluate and compare the performance of optimal p2p codes with the ones optimized for the GIC, and demonstrate that significant improvements are possible.  
We also provide simulation results with specific finite-length codes picked from the optimized code ensembles. 

The rest of the paper is organized as follows. In Section II, system model of the two-user GIC is described. In Section III, HK coding and decoding schemes are explained. In Section IV, the proposed code optimization algorithm is described. In Section V, performance of p2p and optimized LDPC codes are investigated, where finite block length code simulations are also included. Finally, Section VI concludes the paper. 

\section{System Model} 
\begin{figure}[t]
\centering
\includegraphics[scale=.4]{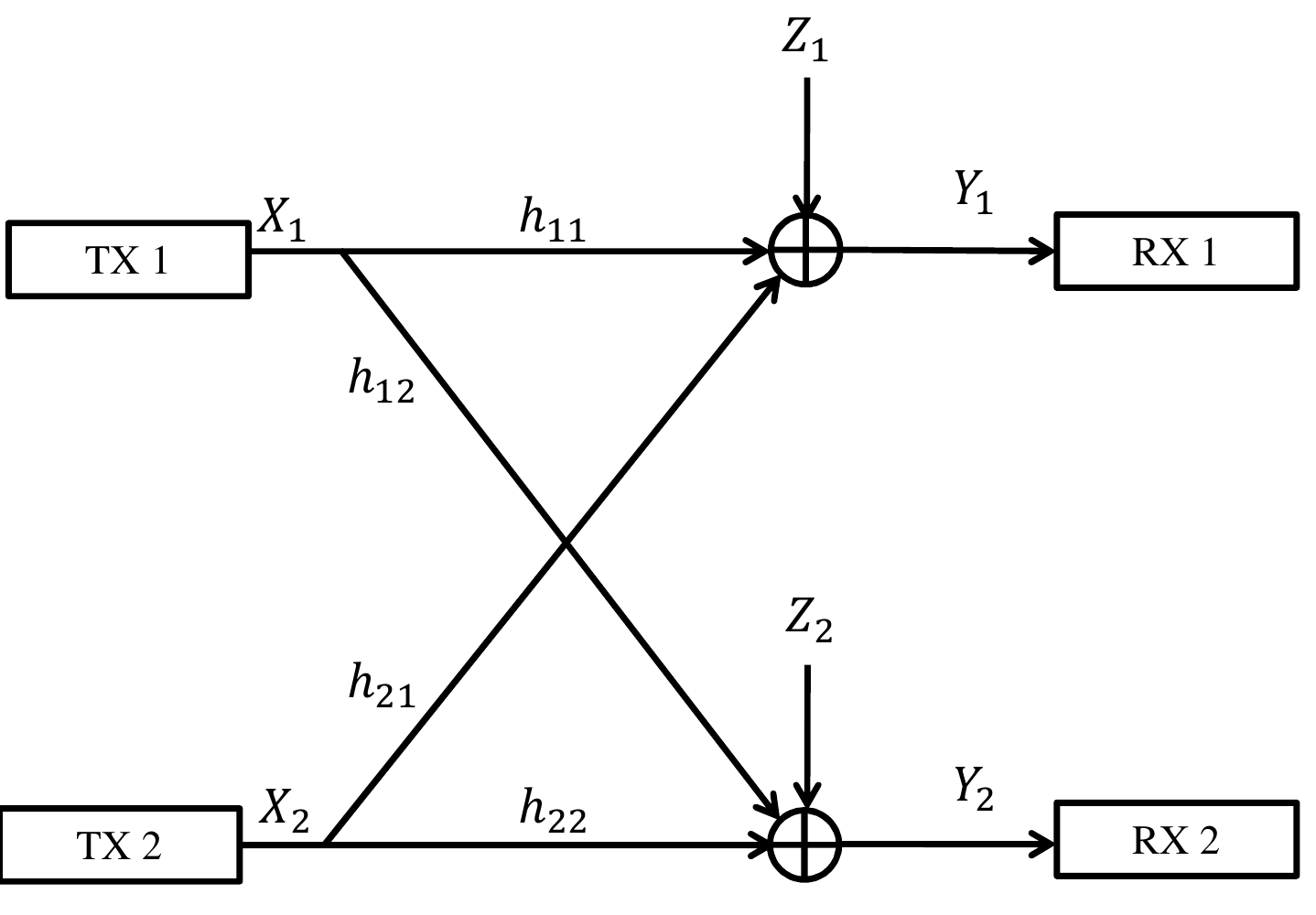}
\caption{Two-user GIC block diagram.}
\label{IC}\vspace{-0.3cm}
\end{figure}

Discrete time two-user GIC system model is illustrated in Fig.~\ref{IC}. Channel outputs at the two receivers can be written as
\begin{equation*}
\begin{array}{l}
Y_1=h_{11}X_1+h_{21}X_2+Z_1,\\
Y_2=h_{12}X_1+h_{22}X_2+Z_2,
\label{GICmodel}
\end{array}
\end{equation*}
where $h_{ij}$ is the channel gain from the user $i$ to the receiver $j$. $Z_1$ and $Z_2$ are independent and identically distributed ~(i.i.d.) circularly symmetric Gaussian noise samples with zero mean and $\frac{N_0}{2}$ variance per dimension. Under per user power constraint, $X_1$ and $X_2$ are the transmitted signals with individual power constraints of $P_1$ and $P_2$, respectively; that is, $E\{|X_i|^2\} \leq P_i$~($i=1,2$). For the case with a total power constraint, the transmitted signals satisfy
\begin{equation*}
\frac{1}{n} \mathop{\sum}_{k=1}^n(P_{1,k}+P_{2,k})\leq P_1+P_2,
\end{equation*} 
where $k$ and $n$ are the index of the transmitted bit and transmission length, respectively. Note that under the above per user power constraint, only a naive TS is possible while with the total power constraint we can employ non-naive TS~\cite{ElGamal2011}. Signal-to-noise-ratios~(SNRs) and interference-to-noise-ratios~(INRs) at receiver $i$ are defined as
\begin{equation*}
SNR_i=\frac{|h_{ii}|^2P_i}{N_0},~INR_i=\frac{|h_{ji}|^2P_j}{N_0},~i,j=1,2,~~i\neq j.
\end{equation*}
Based on the interference to signal level~($a_i=\frac{INR_i}{SNR_i}$) at the receivers, GICs can be categorized as strong ($a_i>1$), weak~($a_i<1$), and mixed~($a_i>1$, $a_j<1$) GICs, where $i\neq j$ and $i,j=1,2$. 
For the case of a symmetric GIC, we have
\begin{equation*}
\begin{array}{c}
h_{11}=h_{22},~~h_{12}=h_{21},\\
SNR_1=SNR_2=SNR,\\
INR_1=INR_2=INR.
\end{array}
\end{equation*}
\vspace{-0.55cm}
\section{Coding and Decoding Schemes} 

\subsection{Encoding}
\begin{figure}[h]
\centering
\includegraphics[scale=.5]{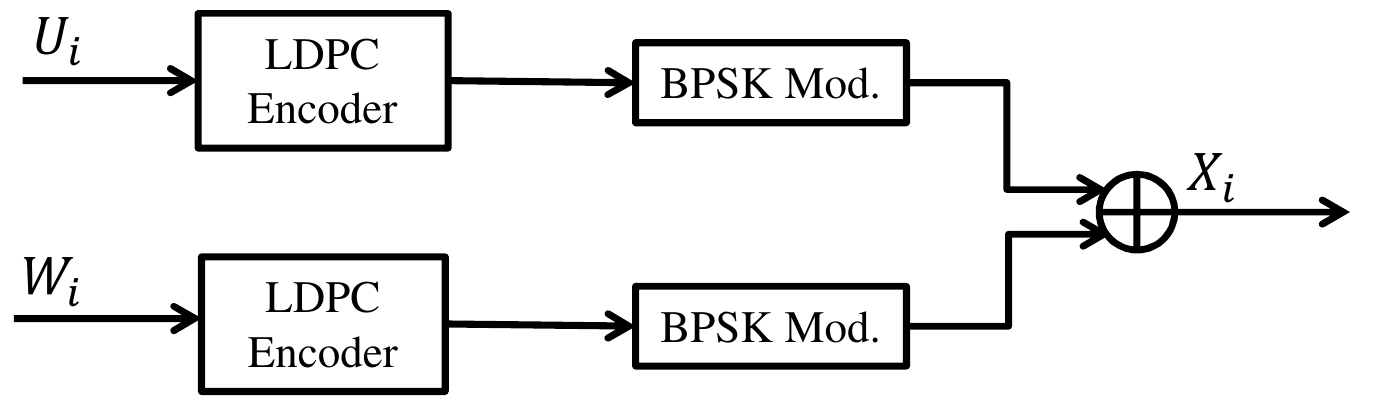}\vspace{-0.25cm}
\caption{Generation of the transmitted signal in the HK coding scheme.}
\label{HKscheme}
\end{figure}

Fig.~\ref{HKscheme} shows the block diagram of the transmitter incorporating the HK coding scheme. As shown in the figure, message of each transmitter is split into public~($W$) and private~($U$) parts, encoded with separate LDPC codes. The encoded private and public messages are then modulated and superimposed at user $i$ with powers $\alpha P_i$ and $(1-\alpha)P_i$, respectively, to form the overall signal to be transmitted, that is,
\begin{equation*}
X_i=\sqrt{\alpha_iP_i}(1-2c_{u_i})+\sqrt{(1-\alpha_i)P_i}(1-2c_{w_i}),~~i=1,2,
\end{equation*}
where $c_{w_i}$ and $c_{u_i}$ are coded bits of public and private messages of user $i$, respectively. In this paper we superimpose two signals with standard addition; however, it is also possible to consider other alternatives. For instance, superimposing of two signals can be done in the ``code'' domain (which may be the proper choice in the case of binary input channels). 

\vspace{-0.2cm}
\subsection{Decoding}
\begin{figure}
\centering
\includegraphics[scale=0.6]{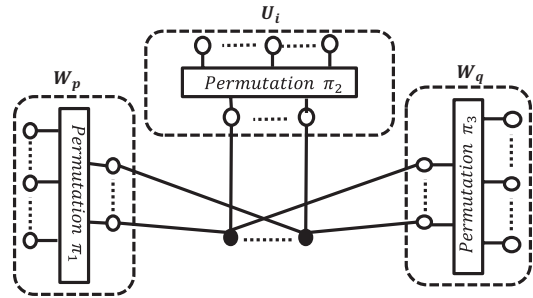}
\caption{JD block diagram~($i,p,q=1,2,~p\neq q$).}
\label{fig:subfig1}
\end{figure}

At the receiver side, the public messages and the private message of the desired user are decoded employing joint decoding~(JD) with parallel scheduling~\cite{Roumy2007}, as illustrated in Fig.~\ref{fig:subfig1}. Under parallel JD, decoding of the messages are performed concurrently and in rounds. Each round starts with computing log-likelihood ratios~(LLRs) fed to the individual decoders, where each decoder runs for several iterations utilizing belief propagation~(BP) algorithm~\cite{RichardsonBook}. Updated LLRs are then passed from variable nodes to intermediate nodes called \emph{state} nodes which completes the round. 
Under BP, LLR of the $i$th coded bit of message $j$, denoted as $c^j_i$, is defined as
\vspace{-0.2cm}
\begin{equation*}
L(c^j_i)=\log\left(\frac{P(c^j_i=0|y)}{P(c^j_i=1|y)}\right).
\label{JD}\vspace{-0.2cm}
\end{equation*}  
Considering parallel scheduling, upon start of each iteration, LLR of the $i$th coded bit provided to individual decoder $j$ is computed at the state nodes by marginalization as follows
\begin{equation*}
L(c^j_i)=\log\left(\frac{\sum_{A_i\in S^{j+}_i}p(y|A_i)p(A_i)}{\sum_{A_i\in S^{j-}_i}p(y|A_i)p(A_i)}\right),
\end{equation*}
where $A_i$ is the vector comprising of the $i$th coded bits of all the (public and private) codewords, i.e.,

\begin{equation*}
A_i=\{x_{u_{1_i}},x_{w_{1_i}},x_{u_{2_i}},x_{w_{2_i}}\}. 
\end{equation*}
$S^{j+}_i$ and $S^{j-}_i$ denote the subset of the $i$th bit of codewords corresponding to $c^j_i=0$ and $c^j_i=1$, respectively. $P(A_i)$ is the probability of the vector $A_i$ whose value is determined by individual decoder outputs and gets updated at each iteration. Considering the $j$th receiver, $u_k$ ($k \ne j$) is not decoded, hence, the corresponding component in $P(A_i)$ does not get updated and remains constant throughout the iterations. 

Symmetry property is defined for the exchanged information (log-likelihood ratio, $L$) in iterative decoding as follows 
\begin{equation}
L=\ln\frac{p(L)}{p(-L)}.
\label{symmetry_condition}
\end{equation}
Here, we state following theorem without proof~(see~\cite{Sharifi2014} for the proof) for all exchanged LLRs under joint decoding adopting parallel scheduling.

\begin{thm}
For a binary-input memoryless output symmetric channel, the probability-density-function~(PDF) of the LLRs exchanged within the factor graph of a joint decoder with parallel scheduling enjoys the symmetry condition in~(\ref{symmetry_condition}).
\end{thm}

\section{LDPC Code Optimization} 

\subsection{Preliminaries}
The objective in this section is to develop an optimization method for LDPC code ensembles over GICs. In this paper, irregular LDPC codes are adopted for transmission. Following the notation in~\cite{Richardson2001a}, an ensemble of irregular LDPC codes~$(\lambda,\rho)$ is described with $\lambda(x)$ and $\rho(x)$, defined as
\begin{equation*}
\lambda(x)=\sum_{i=2}^{d_v}\lambda_i x^{i-1}~~~~~\text{and}~~~~~~\rho(x)=\sum_{i=2}^{d_c}\rho_i x^{i-1}
\end{equation*}
where $d_v$ and $d_c$ are maximum degrees of variable and check nodes, respectively, and the \emph{design rate} of the LDPC code is given by
\begin{equation*}
r=1-\frac{\sum_i \rho_i/i}{\sum_i \lambda_i/i}.
\label{rate}
\end{equation*}

Density evolution is the most accurate available tool to calculate an ensemble's decoding threshold, which tracks the PDF of the exchanged LLRs between variable and check nodes analytically. The difficulty with this method in our case is that under joint decoding, non-linearity of update rule at the state node makes the task of obtaining the PDFs extremely difficult. EXIT~(Extrinsic Information Transfer) chart analysis is an alternate method which tracks the mutual information evolution between the LLRs and transmitted bits under Gaussian assumption for the PDFs. However, in our work, we observed that this method fails for GICs adopting joint decoding approach. In other words, for some ranges of channel parameters, there is a big gap between the thresholds obtained with Gaussian assumption and ones observed through finite block length code simulations, that is why, we do not use this assumption in our analysis.
 
In this paper, mutual information evolution is tracked without any Gaussian assumption on the  exchanged LLRs. To this end, we benefit from Monte-Carlo simulations and run the decoders for a large number of LLR realizations. Armed with symmetry property~\eqref{symmetry_condition}, it is easy to show that the mutual information between the exchanged information in the joint decoder and transmitted bits can be calculated as~\cite{Hagenauer2004}
\begin{eqnarray*}
I(L;X)\hspace{-0.25cm}&=&\hspace{-0.25cm}1-E\{\log_2(1+e^{-L})\}\\
\hspace{-0.25cm}&\approx&\hspace{-0.25cm}1-\frac{1}{N}\sum_{n=1}^N \log_2(1+e^{-x_n.L_n}),
\label{mutualinf}
\end{eqnarray*}
where $x_n$ is the $n$th transmitted symbol. In other words, the mutual information can be tracked and obtained readily without requiring the analytical PDFs of the LLRs. 
\subsection{Proposed Code Optimization Method}
LDPC code design is an optimization problem with nonlinear constraints in general and can be cast using different cost functions such as the threshold or rate maximization. In this paper, we opt for rate maximization under specified channel parameters. The proposed method can be considered as a simple implementation of differential evolution~(DE)\cite{Storn1997} which is a heuristic approach successfully adopted for code optimization over various channels in the previous literature~\cite{Richardson2001a}. The proposed optimization algorithm is based on a \emph{random perturbation technique}~\cite{Franceschini2005} starting with an initial \emph{admissible} $(\lambda,\rho)$ pair~\cite{Bennatan2013}. Without loss of generality and to simplify the exposition, it is assumed that check node degree distribution is a singleton throughout the paper, i.e. $\rho(x)=x^{d_c-1}$, where $d_c$ is determined via an exhaustive search. At each iteration, a perturbation vector {\bf e} is added to the initial variable node degree distribution~($\lambda(x)$).
In order to have a valid distribution, the following constraints should be met
\begin{equation}
\sum_i \lambda_i+e_i=1,
\label{pert1}
\end{equation}
\begin{equation}
0\leq \lambda_i+e_i\leq1,~~~2\leq i\leq d_v,
\label{pert2}
\end{equation}
To satisfy a predetermined increase~($\Delta$) in the code rate, which is a design parameter, we have
\begin{equation*}
1-\frac{1}{d_c}\frac{1}{\sum_i \frac{\lambda_i+e_i}{i}}=r_0+\Delta 
\end{equation*}
\noindent which implies
\begin{equation}
\sum_i \frac{e_i}{i}=\frac{1}{d_c}\frac{\Delta}{(1-r_0)^2-\Delta(1-r_0)}.
\label{pert5}
\end{equation}
In other words, the perturbation vector should satisfy \eqref{pert1},~\eqref{pert2},~\eqref{pert5}. The new degree distribution will replace the initial degree distribution if it is admissible, else it is dismissed and a new iteration is performed. The process is stopped if a new admissible degree distribution cannot be found after a predetermined number of iterations. 
\section{Examples of LDPC Codes Over GICs}
In this section, we investigate the performance of irregular LDPC codes adopted for transmission over the two-user GIC implementing the HK coding/decoding strategy. Two scenarios with BPSK modulation and fixed channel gains are considered. In the examples, we optimize the LDPC codes jointly with the goal of maximizing the sum rate. The achieved rate pairs are then compared to the best achievable rate pairs employing optimized binary-input additive-white-Gaussian-noise~(BI-AWGN) codes. EXIT chart analysis~\cite{Brink2004} is utilized to optimize the degree distributions for BI-AWGN codes (i.e., to generate the p2p codes to provide a benchmark). To perform the optimization, we consider a singleton check node degree distribution which is determined via an exhaustive search. Inspired by~\cite{Richardson2001a}, nonzero variable node degrees are limited to \{2,3,4\}, the maximum degree $d_v$~($50$), and a few degrees in-between. Specifically, in our paper we select these degrees as \{2,3,4,9,10,19,20,49,50\}.
\vspace{-0.1cm}
\subsection{Scenario~I~--~Strong GIC}
\begin{figure}[ht]
\centering
\includegraphics[scale=0.7]{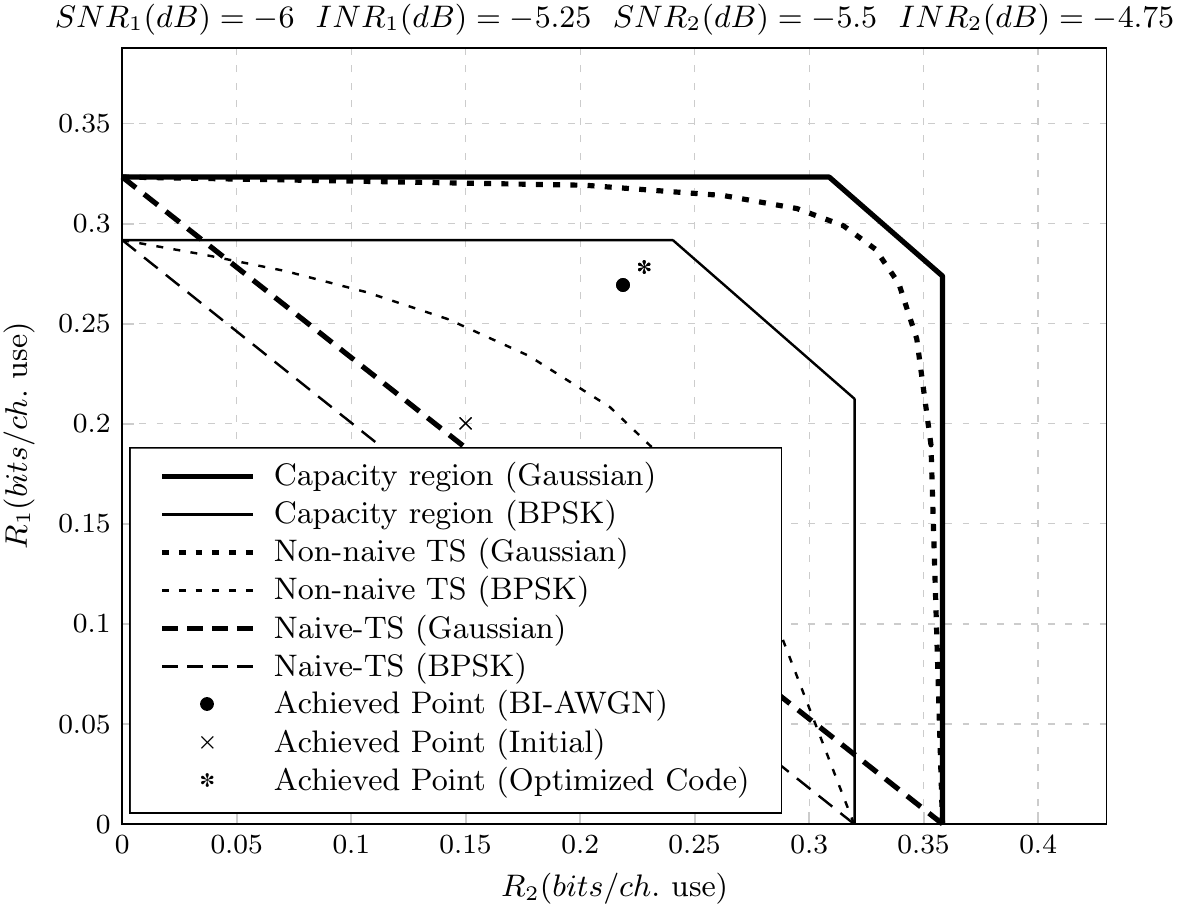}\vspace{-0.3cm}
\caption{Scenario~I:~regions and achieved points for strong GIC.}\vspace{-0.3cm}
\label{fig:bpsk-str-asym}\vspace{-0.15cm}
\end{figure}

In this example, all messages are public since the interfering signal is strong and can be decoded completely at the receiver side. we consider an asymmetric rate pair whose sum rate is maximized with the following constraint
\vspace{-0.2cm}
\begin{equation*}
R_1=R_2+K, \vspace{-0.1cm}
\label{const:1}
\end{equation*}
where $K$ is a constant. The channel parameters are given in Fig.~\ref{fig:bpsk-str-asym}. Specifically, we aim to get close to one of the corner points of the capacity region. As a result, $K$ is set to $0.05$ and fixed throughout the code optimization procedure. The algorithm is initialized with the rate pair~($0.2,0.15$). Optimized degree distributions are given in Table~\ref{tab:bpsk-str-asym}, where superscripts $P$ and $O$ refer to p2p and optimized degree profiles, respectively. It can be observed that the optimized achieved rate pair (0.278, 0.228) is superior to the rate pair (0.269, 0.219) obtained by the best p2p codes. In addition, the achieved rate pair outperforms even the non-naive time-sharing scheme. It is also noted that the achieved point is $0.31$~dB away from the capacity boundary corresponding to BPSK signaling~(estimated by increasing the noise variance and maintaining the $\frac{INR}{SNR}$ ratio).

\begin{table*}[ht]
\center
\caption{Optimized Degree Distributions~($\rho_{W_{1_P}}(x)=x^4$, $\rho_{W_{1_O}}(x)=x^4$, $\rho_{W_{2_P}}(x)=x^4$, $\rho_{W_{2_O}}(x)=x^3$).}\vspace{-0.25cm}
\begin{tabular}{|c|c|c|c|c|c|c|c|c|c|}
\hline & $\lambda_2$ & $\lambda_3$ & $\lambda_4$ & $\lambda_9$ & $\lambda_{10}$ & $\lambda_{19}$ & $\lambda_{20}$ & $\lambda_{49}$ & $\lambda_{50}$\\
\hline 
$W_{1_P}$ & 0.2759 & 0.2502 & 0.1001 & 0.1089 & 0.0502 & 0.1706 & 0.0086 & 0.0247 & 0.0108 \\ \hline
$W_{2_P}$ & 0.2575 & 0.2490 & 0.0619 & 0.1320 & 0.0768 & 0.0586 & 0.0037 & 0.0494 & 0.1111 \\
\hline 
$W_{1_O}$ & 0.3106 & 0.1901 & 0.1065 & 0.1691 & 0.0809 & 0.0337 & 0.0297 & 0.0033 & 0.0761\\
\hline
$W_{2_O}$ & 0.3815 & 0.2999 & 0.0280 & 0.1453 & 0.0719 & 0.0340 & 0.0074 & 0.0093 & 0.0227 \\
\hline  
\end{tabular}\vspace{-0.25cm}
\label{tab:bpsk-str-asym}
\end{table*}
\vspace{0.1cm}
\subsection{Scenario~II~--~Weak GIC}
\begin{figure}[ht]
\centering
\includegraphics[scale=0.7]{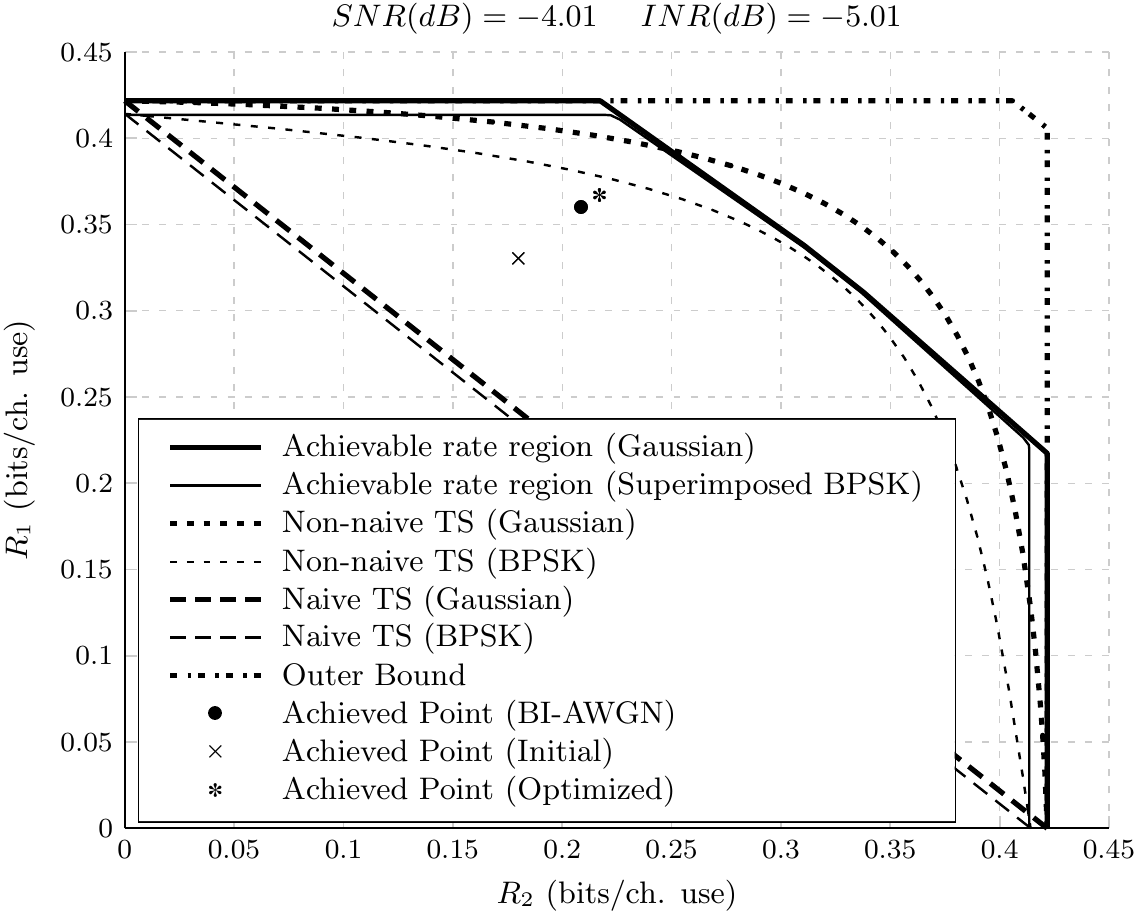}
\caption{Scenario~III:~regions and achieved points for weak GIC.}
\label{ARR5}\vspace{-0.15cm}
\end{figure}

For this instance, a symmetric weak GIC is considered for which the channel parameters are as given in Fig.~\ref{ARR5}. The signaling is named as superimposed BPSK since the transmitted signal for each user is a combination of public and private messages. Since interfering signals cannot be decoded completely, portion of the power is allocated to private messages. To address the power allocation problem, the following optimization is considered
\vspace{-0.1cm}
\begin{equation}
\begin{array}{rl}
\underset{\alpha_1, \alpha_2} {\max} &R_1+R_2  \\
\text{subject to}&R_1=R_2+K,\\
&0\leq\alpha_i \leq 1,~~i=1,2,
\end{array}
\label{opt}
\end{equation}
where $K$ is set to $0.15$.
It should be noted that in solving the optimization~\eqref{opt}, individual rate~(private and public) constraints should be satisfied~\cite[(3.2)-(3.15)]{Kobayashi1981}. Having obtained power and rate assignments, three LDPC code ensembles (i.e., for encoding $U_1$, $W_1$, and $W_2$ considering receiver 1) are optimized. The initial rates used in code optimization are as follows: $R_{U_1}=0.125, R_{W_1}=0.205, R_{W_2}=0.18$. The resulting degree distributions are given in Table~\ref{tab:bpsk-pp-p-optimized}. It can be shown that the optimized achieved rate pair (0.367,0.217) is higher than the rate pair (0.36, 0.209) obtained by the best p2p codes. Further, the achieved rate pair is better than the one obtained with the single user code utilizing naive TS. It is also clear that optimized degree distributions exhibit better performance than optimized p2p pairs. 
\begin{table*}[ht]
\centering
\caption{Optimized Degree Distributions ($\rho_{U_{1_P}}(x)=\rho_{U_{1_O}}(x)=x^4, \rho_{W_{1_P}}(x)=\rho_{W_{1_O}}(x)=x^3, \rho_{W_{2_P}}(x)=x^4, \rho_{W_{2_O}}(x)=x^3$).}\vspace{-0.25cm}
\begin{tabular}{|c|c|c|c|c|c|c|c|c|c|}
\hline & $\lambda_2$ & $\lambda_3$ & $\lambda_4$ & $\lambda_9$ & $\lambda_{10}$ & $\lambda_{19}$ & $\lambda_{20}$ & $\lambda_{49}$ & $\lambda_{50}$\\
\hline 
$U_{1_P}$ & 0.2659 & 0.2455 & 0.0512 & 0.1661 & 0.0542 & 0.0203 & 0.0415 & 0.0546 & 0.1007 \\ \hline
$W_{1_P}$ & 0.3488 & 0.1237 & 0.2267 & 0.0161 & 0.0912 & 0.0299 & 0.0422 & 0.0971 & 0.0243\\ \hline
$W_{2_P}$ & 0.2386 & 0.2859 & 0.0504 & 0.0920 & 0.0892 & 0.0326 & 0.0176 & 0.1183 & 0.0754\\ \hline
$U_{1_O}$ & 0.3151 & 0.0566 & 0.2480 & 0.0523 & 0.0213 & 0.1759 & 0.0288 & 0.0405 & 0.0615 \\ \hline
$W_{1_O}$ & 0.3642 & 0.1226 & 0.2030 & 0.0222 & 0.0460 & 0.0163 & 0.1385 & 0.0370 & 0.0502 \\ \hline
$W_{2_O}$ & 0.4309 & 0.1642 & 0.1127 & 0.0969 & 0.0481 & 0.0384 & 0.0404 & 0.0578 & 0.0106 \\
\hline
\end{tabular}\vspace{-0.35cm}
\label{tab:bpsk-pp-p-optimized}
\end{table*}

\subsection{Finite Block Length Code Simulations}
\begin{figure}
\centering
\includegraphics[scale=0.72]{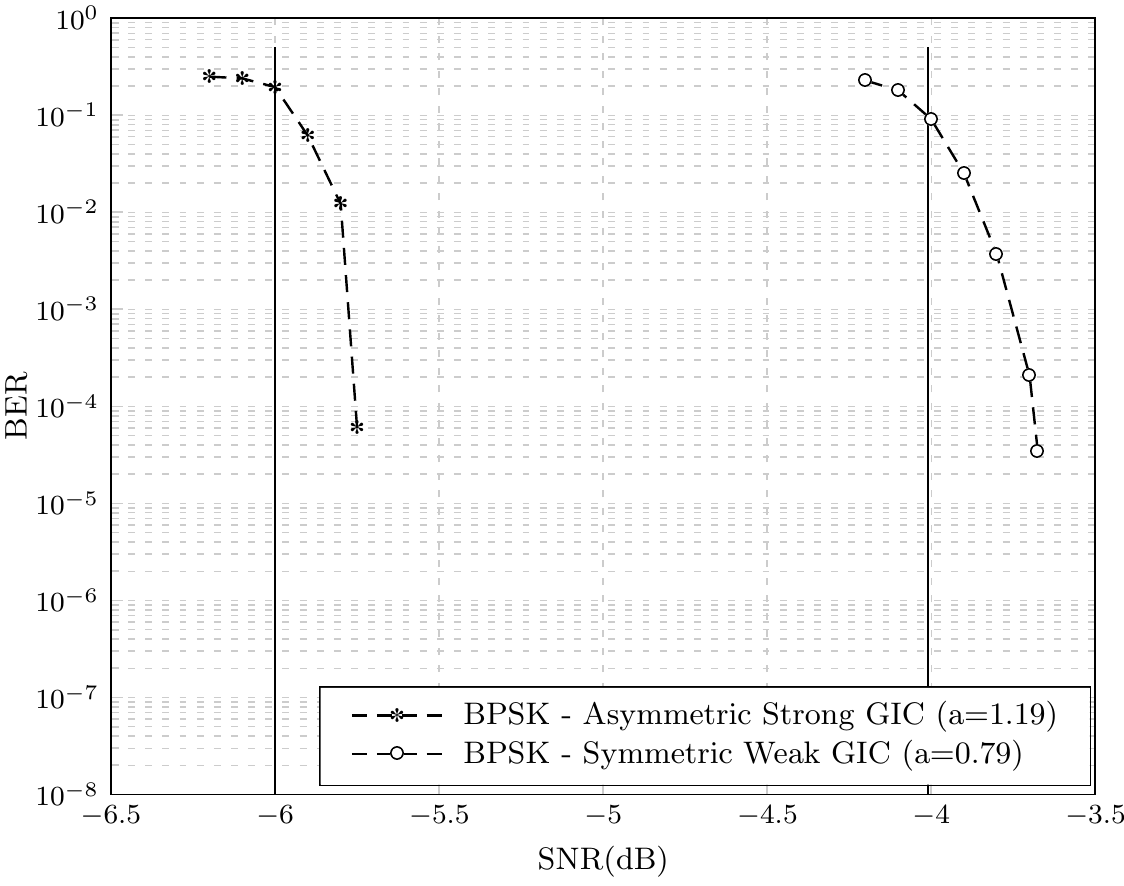}\vspace{-0.3cm}
\caption{Finite block length code decoding results.}\vspace{-0.48cm}
\label{fig:BER}
\end{figure}

In this section, we evaluate the performance of the optimized LDPC codes through finite block length simulations. We consider codes with block lengths $50$k and set the maximum number of decoding iterations to $500$. Fig.~\ref{fig:BER} shows the decoding results at receiver 1. For clarity of presentation, we include decoding result of the public or private messages with the worst error rates (i.e., the bottleneck), instead of giving the results of all the cases (could be up to six different decoding results in the general case). Considering bit-error-rate~(BER) of $10^{-4}$ as the reliable transmission, it can be observed that decoding results are within $0.3$~dB of the decoding thresholds.

\section{Conclusions}

In this paper, the Han-Kobayashi~(HK) coding/decoding strategy is implemented for the two-user Gaussian interference channel~(GIC) considering fixed channel gains and finite constellations. A robust method is proposed for LDPC code optimization utilizing a random perturbation method. Performance of the designed LDPC codes are examined for strong and weak interference levels through examples. Under strong GIC, capacity approaching codes are designed which beat even the non-naive TS rate region. Under weak interference, it is observed that optimized codes beat the naive TS region (with Gaussian signaling) and operate close to the non-naive TS region boundary. We also note that the designed codes improve consistently on the codes designed for p2p channels (used with the same encoding/decoding procedure). Simulation results are also provided using codes picked from the designed LDPC code ensembles to verify the asymptotic results. 

\section*{Acknowledgment}

Part of this publication, specifically Section 4, was made possible by NPRP grant 4-1293-5-213 from the Qatar National Research Fund (a member of Qatar Foundation). In addition, we also acknowledge support from National Science Foundation under the grant NSF-CCF 1117174 for Sections 1-3,5,6. The statements made herein are solely the responsibility of the authors.

\bibliographystyle{IEEEtran}
\bibliography{bib_new}

\begin{thebibliography}{10}
\providecommand{\url}[1]{#1}
\csname url@samestyle\endcsname
\providecommand{\newblock}{\relax}
\providecommand{\bibinfo}[2]{#2}
\providecommand{\BIBentrySTDinterwordspacing}{\spaceskip=0pt\relax}
\providecommand{\BIBentryALTinterwordstretchfactor}{4}
\providecommand{\BIBentryALTinterwordspacing}{\spaceskip=\fontdimen2\font plus
\BIBentryALTinterwordstretchfactor\fontdimen3\font minus
  \fontdimen4\font\relax}
\providecommand{\BIBforeignlanguage}[2]{{%
\expandafter\ifx\csname l@#1\endcsname\relax
\typeout{** WARNING: IEEEtran.bst: No hyphenation pattern has been}%
\typeout{** loaded for the language `#1'. Using the pattern for}%
\typeout{** the default language instead.}%
\else
\language=\csname l@#1\endcsname
\fi
#2}}
\providecommand{\BIBdecl}{\relax}
\BIBdecl

\bibitem{Kobayashi1981}
T.~Han and K.~Kobayashi, ``{A new achievable rate region for the interference
  channel},'' \emph{IEEE Transactions on Information Theory}, vol.~27, no.~1,
  pp. 49--60, Jan. 1981.

\bibitem{Roumy2007}
A.~Roumy and D.~Declercq, ``{Characterization and optimization of LDPC codes
  for the 2-user Gaussian multiple access channel},'' \emph{EURASIP Journal on
  Wireless Communication and Networking}, {Article ID: 74890, 2007}.

\bibitem{Berlin2005}
P.~Berlin and D.~Tuninetti, ``{LDPC codes for fading Gaussian broadcast
  channels},'' \emph{IEEE Transactions on Information Theory}, vol.~51, no.~6,
  pp. 2173--2182, Jun. 2005.

\bibitem{Hu2007}
J.~Hu and T.~M. Duman, ``{Low density parity check codes over wireless relay
  channels},'' \emph{IEEE Transactions on Wireless Communications}, vol.~6,
  no.~9, pp. 3384--3394, Sep. 2007.

\bibitem{Bennatan2013}
A.~Bennatan, S.~Shamai~(Shitz), and A.~R. Calderbank, ``{Soft-decoding-based
  strategies for relay and interference channels: analysis and achievable rates
  using LDPC codes},'' Mar. 2013, arXiv:1008.1766.

\bibitem{Storn1997}
R.~Storn and K.~Price, ``{Differential evolution--A simple and efficient
  heuristic for global optimization over continuous spaces},'' \emph{Journal of
  Global Optimization}, vol.~11, pp. 341--359, Dec. 1997.

\bibitem{Sharifi2014}
S.~Sharifi, A.~K. Tanc, and T.~M. Duman, ``{On LDPC Codes for Interference
  Channels},'' \emph{Submitted to IEEE Transactions on Communications}.

\bibitem{ElGamal2011}
A.~{El Gamal} and K.~Young-Han, \emph{{Network Information Theory}}.\hskip 1em
  plus 0.5em minus 0.4em\relax {Cambridge University Press}, 2011.

\bibitem{RichardsonBook}
T.~Richardson and R.~Urbanke, \emph{Modern Coding Theory}.\hskip 1em plus 0.5em
  minus 0.4em\relax Cambridge University Press, 2008.

\bibitem{Richardson2001a}
T.~Richardson, M.~Shokrollahi, and R.~Urbanke, ``{Design of
  capacity-approaching irregular low-density parity-check codes},'' \emph{IEEE
  Transactions on Information Theory}, vol.~47, no.~2, pp. 619--637, Feb. 2001.

\bibitem{Hagenauer2004}
J.~Hagenauer, ``{The EXIT chart--introduction to extrinsic information transfer
  in iterative processing},'' \emph{in 12th European Signal Processing
  Conference}, 2004.

\bibitem{Franceschini2005}
M.~Franceschini, G.~Ferrari, R.~Raheli, and A.~Curtoni, ``{Serial concatenation
  of LDPC codes and differential modulations},'' \emph{IEEE Journal on Selected
  Areas in Communications}, vol.~23, no.~9, pp. 1758--1768, Sep. 2005.

\bibitem{Brink2004}
S.~ten Brink, G.~Kramer, and A.~Ashikhmin, ``{Design of low-density
  parity-check codes for modulation and detection},'' \emph{IEEE Transactions
  on Communications}, vol.~52, no.~4, pp. 670--678, Apr. 2004.

\end{thebibliography}

\end{document}